\documentclass[multphys,vecphys]{svmult}

\usepackage{makeidx}         
\usepackage{graphicx}        
\usepackage{multicol}        
\usepackage[bottom]{footmisc}

\makeindex             

\begin{document}

\title*{The lifecycle of powerful AGN outflows}
\author{Christian R. Kaiser\inst{1}\and
Philip N. Best\inst{2}}
\institute{School of Physics \& Astronomy, University of Southampton, Southampton, SO17 1BJ, UK \texttt{crk@soton.ac.uk}
\and Institute for Astronomy, Royal Observatory Edinburgh, Blackford Hill, Edinburgh, EH9 3HJ, UK \texttt{pnb@roe.ac.uk}}
%
%
\maketitle

\section{Introduction}

During the course of this conference, much evidence was presented that points to an intimate connection between the energetic outflows driven by AGN and the energy budget and quite possibly also the evolution of their gaseous environments. However, it is still not clear if and how the AGN activity is triggered by the cooling gas, how long the activity lasts for and how these effects give rise to the observed distribution of morphologies of the outflows. 

In this contribution we concentrate on the high radio luminosity end of the AGN population. While most of the heating of the environmental gas may be due to less luminous and energetic outflows \cite{bkh06}, these more powerful objects have a very profound influence on their surroundings \cite{ba02}. We will describe a simple model for powerful radio galaxies and radio-loud quasars that explains the dichotomy of their large-scale radio morphologies as well as their radio luminosity function. 

\section{The FR dichotomy}

In their original paper \cite{fr74} Fanaroff and Riley pointed out that the large-scale radio structure of radio-loud AGN is either edge-darkened (FRI) or edge-brightened (FRII) and that there is a reasonably sharp transition between these two morphologies at around a luminosity of $10^{26}$\,W\,Hz$^{-1}$ at an observing frequency of 151\,MHz. High-resolution radio observations show thin, presumably laminar jet flows ending in strong shocks for the more luminous FRII-type objects. After passing through the shock, visible as luminous `hotspots' in radio observations, the jet material inflates radio lobes around the jets. Objects classed as FRI-types show a more varied radio morphology. While some show radio lobes reminiscent of those of FRII-type object without radio hotspots, other FRI-type objects contain turbulent jets. The appearance of these turbulent jets is similar to that of smoke rising out of a chimney. Here we concentrate on the latter class of objects.

There are also several other observational differences between the AGN giving rise to the jet flows in the two FR classes. Usually FRI-type objects show weaker optical continuum \cite{ccc00} and line emission \cite{hl79}. Their X-ray emission is also less luminous \cite{hec06}. All this evidence points to weaker emission from the accretion disc in the AGN in FRI-type sources, which is also consistent with less radiative heating of the dust in their host galaxies \cite{mhs04}. However, the so-called Low-Excitation Radio Galaxies (LERG) with FRII morphology and radio luminosities close to the value separating the FR classes appear to contain AGN more similar in their emission properties to those usually found in FRI-type objects \cite{ccc00,bzo95,hec06}. The opposite combination of a bright AGN with an FRI-type radio morphology appears to be rare, but some detections have been claimed \cite{br01}. 

Finally, the dividing line between the FR classes in terms of radio luminosity is related to the optical luminosity of the host galaxy as $L_{\rm B}^{1.8}$ in the B-band \cite{lo96}.

\section{The Radio Luminosity Function (RLF)}

The RLF at low cosmological redshifts is well approximated by a broken power-law \cite{dp90}. The cosmological evolution of the radio luminosity function seems to imply that this shape is caused by two distinct populations, one of which dominates below the break and the other contributing mainly at luminosities above the break \cite{dp90,wrb01}. The break luminosity of the RLF is suspiciously close to the dividing line between the two FR classes. However, current samples of extragalactic radio sources do not contain information on radio morphology for enough sources or are too small to allow the determination of separate RLFs for the two classes. 

In the following we will use a break luminosity of $P_{\rm break} \sim 5 \times 10^{25}$\,W\,Hz$^{-1}$ at an observing frequency of 1.4\,GHz. We also use the power-law slopes of the RLF as $-0.6$ below $P_{\rm break}$ and $-2.4$ above the break \cite{wrb01}. Note here, that observationally the RLF is determined in units of density per logarithm of radio luminosity. If we work in units of density per radio luminosity, then the slopes of the RLF are $-1.6$ and $-3.4$, respectively.

\section{A unified model for both FR classes and the RLF}

\subsection{Modelling FRI-type sources}

Due to their turbulent jets, it is notoriously difficult to construct models for the evolution of FRI-type sources. The flow structure is complicated and virtually impossible to model accurately. This implies that the strength and structure of the magnetic field is not predictable. It is also not clear how and where the relativistic electrons giving rise to the observed synchrotron radiation are accelerated. Hence it is also not possible to estimate how much of the energy transported by the jets is dissipated to these electrons. 

Some insight can be gained from arguments based on conservation laws \cite{gb95}, but this does not lead to a straightforward model describing the evolution of the jet flow and the emitted radiation. Models of the velocity field of the jet and the underlying magnetic fields and distributions of relativistic electrons can be crafted on high resolution radio images of FRI-type sources \cite{lb04}, but these models do not make predictions for the emission properties of these objects in general. However, both approaches can be used as guides towards some rough principles determining the radio luminosity and its evolution. 

Most FRI-type source with turbulent jets show a low luminosity, confined jet similar to the jets seen in FRII-type morphologies extending for a short distance from the radio core. At the end of this possibly laminar jet the flow suddenly widens and brightens considerably in the `flare point'. After this the jet appears to be completely turbulent, it spreads further and fades into the surroundings as it travels outwards. The sudden increase in the luminosity of the jet in the flare point and the fading of the emission further out may indicate that most of the emitting particles are accelerated in the flare point with little or no additional acceleration at larger distances. If the energy transport rate of the jet or `jet power' and the flare point are not evolving, then a zeroth-order model for FRI sources would predict their radio luminosity to remain constant throughout their lifetime. Furthermore, if the flare point does not evolve, then it must be in pressure equilibrium with the surrounding gas. It is then very likely that the radio luminosity of the flare point and that of the turbulent jet are linearly proportional to the jet power.

In the following we adopt this simple model for FRI-type sources. If we now assume that the lifetime of the jet flows does not depend on the jet power, $Q$, and also that the RLF is dominated by FRI-type objects below $P_{\rm break}$, then we can infer the `birth rate' of radio-loud AGN directly from the RLF to be proportional to $Q^{-1.6}$. 

\subsection{Modelling FRII-type sources}

It is possible to formulate analytical models for the radio luminosity evolution of FRII-type objects. The reason is that in this class the underlying fluid flow is laminar and hence more amenable to modelling. Also, high resolution radio observations imply little or no acceleration of relativistic particles away from the radio hotspots at the ends of the laminar jets \cite{cpdl91}. 

Most current models are based on the fundamental picture of a jet in pressure equilibrium with its own lobe \cite{ps74}. Once the dynamics of the jet-lobe system are established \cite{ka96b}, we can calculate the resulting radio emission taking into account the cumulative energy losses of the radiating electrons \cite{kda97a}. The gas density in the environment of the source is modelled as a modified $\beta$-profile, i.e.
\begin{equation}
\rho = \frac{\rho_0}{\left[ 1+\left( r /a_0 \right)^2 \right]}^{\beta / 2}.
\end{equation}
This profile is well approximated by a constant density $\rho_0$ inside the core radius $a_0$ and by a simple power-law $\rho = \rho_0 \left( r / a_0 \right)^{-\beta}$ further out. The luminosity evolution in the power-law regime is given by $P \propto t^{-1/2}$ for $\beta = 2$. Here we neglect radiative losses of the relativistic electrons due to the emitted synchrotron radiation as these mainly affect young objects \cite{kda97a}. We also ignore the energy losses due to Inverse Compton (IC) scattering of cosmic microwave background (CMB) photons off the electrons because these affect mainly old sources (see below). 

If we assume that the RLF above $P_{\rm break}$ is dominated by FRII-type objects, then we can determine the slope of the RLF from the birth function of sources with a given jet power and the luminosity evolution determined above. The model predicts an RLF proportional to $P^{-3.4}$ in excellent agreement with the observations.

\subsection{The connection between the FR classes}

What decides the radio morphology a given sources develops? This must clearly be connected to the onset of turbulent disruption of the jet flow. In FRII-type sources the jets are located inside the radio lobes. These contain mainly the material that has been transported along the jet. From geometric considerations it follows that the gas density in the lobe is lower than that inside the jet. Therefore the jets in FRII sources are very well protected against turbulent disruption which would require the jets to be in contact with much denser gas. 

The inflation of lobes depends crucially on the formation of a bow shock around the lobes. In other words, the lobes must be overpressured with respect to the ambient gas. If they are not, then Rayleigh-Taylor instabilities develop on the lobe surface \cite{ps74}. In this case, the denser gas of the source environment will start to buoyantly replace the lobe material until it reaches the jet itself. At this point the interaction of the jet with the dense material will lead to turbulent disruption of the jet. So, the crucial condition for the jet to remain laminar is the overpressured lobe. 

At large distances the pressure in the source environment will decrease proportional to $r^{-2}$. The dynamical model of the jet-lobe system predicts that the pressure inside the lobe also decreases with $r^{-2}$ in this regime \cite{ka96b}. Hence the lobe remains overpressured with respect to its surroundings at all times provided it was overpressured in the beginning. The situation is quite different in the core region, where the external pressure is essential constant, but the pressure inside the lobe decreases as $r^{-4/3}$. Since the pressure inside the lobe also depends on the jet power, we can summarise this result as follows. All sources start out with an FRII-type morphology. The lobe pressure in sources with weaker jets falls so rapidly that they come into pressure equilibrium with their surroundings while they are still located in the core region. Their jets develop turbulent flows with an FRI-type radio morphology. These sources dominate the RLF at low radio luminosities. More powerful jets grow beyond the core region before the pressure in their lobes decreases too much. They retain their FRII-type morphology and form the bulk of the population at the high luminosity end of the RLF. 

For a given environment density distribution we can calculate from the model the jet power at which the divide between the FR classes occurs. For reasonable values of the necessary parameters we obtain $10^{37}$\,W, close to the jet power suggested for the divide from observations of emission line strengths \cite{rs91}. We can also determine the radio luminosity associated with the divide. This depends on the density of the source environment which also determines the X-ray luminosity of this gas. Using the observed correlation of X-ray luminosity and luminosity in the B-band of giant elliptical galaxies \cite{ofp01}, we predict that the dividing line between the FR classes should occur at around $10^{26}$\,W\,Hz$^{-1}$ with a dependence on the B-band luminosity of the host galaxy as $L_{\rm B}^{1.5 \rightarrow 2}$. Both predictions are again in very good agreement with observations \cite{wrb01,lo96}.

\section{A second switch}

The above model establishes a connection between the break in the RLF and the division between the FR classes in terms of radio morphology on the one side and the jet power and the density of the source environment on the other. Therefore in its current form it cannot explain the differences between the FR classes observed outside the radio band.

The observational evidence suggests that the AGN predominantly associated with large-scale radio structure of type FRI are comparatively inefficient at producing observable radiation. For jet producing X-ray binaries the radiatively inefficient accretion state is associated with the production of weak jets. Powerful jet ejections are connected to the transition to the radiatively efficient accretion state \cite{fbg04}. If the same connection applies to the accretion discs of AGN and their jets, then this would explain the observations as less powerful jets arising from radiatively inefficient accretion discs are more susceptible to turbulent disruption and hence preferentially develop FRI-type morphologies. 

As we have seen above, the division between the FR classes also depends  not only on the jet power but also on the properties of the source environment. Hence we may expect to observe some hybrid objects, i.e. radiatively inefficient AGN with an FRII-type radio morphology or bright AGN with FRI-type radio structures. The former can be identified as the LERGs. Their radio luminosities are always close to the dividing line between the FR classes \cite{pb89} and they also often show weak radio hotspots and bright jet flows \cite{hap98}. Both properties may indicate relatively weak jets that are affected by turbulence. 

Radiatively efficient AGN with turbulent jets appear to be rare. This may indicate that the jet power associated with the division of radiatively efficient and inefficient accretion is below $10^{37}$\,W; the jet power at which the FR divide occurs in typical source environments. It will be interesting to further pursue this connection between the jets emerging from AGN and those of X-ray binary stars. In particular the evidence for the X-ray binaries shows that individual objects cycle through accretion states and produce jets of very different power along the way. Is this also true for AGN?

%


\printindex
\end{document}